\documentclass[aps,prl,reprint,groupedaddress, showpacs]{revtex4-1}
\usepackage{graphicx}
\usepackage{natbib}

\begin{document}


\title{Multipixel sub-shot noise phase measurement with classical light interferometry}


\author{Mandeep Singh, Kedar Khare}
\email[]{kedark@physics.iitd.ac.in}
\affiliation{Department of Physics, Indian Institute of Technology Delhi, 110016 INDIA}

\author{Anand Kumar Jha}
\affiliation{Department of Physics, Indian Institute of Technology Kanpur, 208016 INDIA}

\author{Shashi Prabhakar, R. P. Singh}
\affiliation{Physical Research Laboratory, Ahmedabad, 380009 INDIA}

\date{\today}

\begin{abstract}
We demonstrate accurate phase measurement from low photon level interference data using a constrained optimization method that takes into account the expected redundancy in the unknown phase function. This approach is shown to have significant noise advantage over traditional methods such as balanced homodyning or phase shifting that treat individual pixels in the interference data as independent of each other. Our interference experiments comparing the optimization method with the traditional phase shifting method show that when the same photon resources are used, the optimization method provides phase recoveries with tighter error bars. In particular, RMS phase error performance of the optimization method for low photon number data (10 photons per pixel) shows $>$ 5X noise gain over the phase shifting method. In our experiments where a laser light source is used for illumination, the results imply phase measurement with accuracy better than the conventional single pixel based shot noise limit (SNL) that assumes independent phases at individual pixels. The constrained optimization approach presented here is independent of the nature of light source and may further enhance the accuracy of phase detection when a nonclassical light source is used.
\end{abstract}

\pacs{42.30.Rx, 42.25.Hz, 42.50.St, 42.87.Bg, 42.30.-d}

\maketitle


Interferometric phase detection is one of the most important techniques in Physics. Optical interferometers are being used routinely for metrology, biomedical applications, Fourier transform spectroscopy, and holographic 3D imaging to name a few applications \cite{Hariharan2007}. Sensitive phase detection is at the heart of large scale collaborative efforts such as gravitational wave detection \cite{Abbott2009}. Our aim in this Letter is to examine the interferometric phase detection problem with an optimization framework that effectively models the redundancy in the unknown phase signal. For given photon resources, we show that this approach gives phase measurement with accuracy better than the conventional single pixel based shot noise limit (SNL) even when a classical light source is used. This conclusion, though somewhat surprising, suggests that limits such as SNL may be generalized to incorporate the multipixel structure of the unknown phase signal. As discussed later, while the quantum limits to measurement of stochastically fluctuating time-varying phase have been studied before, our focus in this work is to exploit the redundancy in the phase signal to obtain enhanced phase measurement accuracy.   

When two complex fields $R$ (reference field) and $O$ (object field) interfere, the interference signal $I$ detected by a square law detector is represented by:
\begin{equation}\label{eq:interference}
I = |R|^2 + |O|^2 + R^{\ast}O + R O^{\ast}.
\end{equation}   
Given the prior knowledge about $R$, the typical methods for analysis of the interference data are linear in nature.  The first step in estimating phase from interference data is to get rid of the two intensity terms $|R|^2$ and $|O|^2$ in Eq. (\ref{eq:interference}) followed by processing of the remaining cross terms to estimate the amplitude and the phase of the unknown complex field $O$. The removal of $|R|^2$ and $|O|^2$ may be performed by high pass filtering of the interference signal $I$ or by using multiple recordings of the interference signal with known phase shifts in $R$. When the phase of $O$ is smaller than $\pi/2$ in magnitude, a balanced detection scheme such as homodyning \cite{Haus2000} may be followed. However, if the phase of $O$ can take any value in the interval $[-\pi, \pi]$, typically four interference signals are recorded with refernce phase shifts of $\theta = 0$, $\pi/2$, $\pi$, and $3\pi/2$ applied to $R$ \cite{Yamaguchi1997}. The corresponding four interference records are sufficient to provide information about the two quadratures of the unknown object field $O$. Denoting the four interference records as $I_{\theta}$, the phase $\phi_O$ of the object field relative to the phase $\phi_R$ of the reference field may be expressed as:
\begin{equation}\label{eq:PS}
\phi_O - \phi_R = \arctan\Big(\frac{I_{3\pi/2} - I_{\pi/2}}{I_{0} - I_{\pi}}\Big).
\end{equation}
Henceforth, we will refer to this procedure as the phase shifting method (PSM). Improving the accuracy of the phase estimation is of great interest to all the associated applications and this problem has been studied in detail in literature \cite{{Walkup1973}, {Caves1981}, {Jaekel1990}, {Depeursinge2007}}. It is now well established that when classical light sources are used, the phase detection accuracy is ultimately limited by the shot noise or the $\sqrt{N}$ noise where $N$ is the mean number of photon counts registered by a point detector. This noise limit is often referred to as the SNL. Obtaining phase detection accuracy below the SNL requires the use of non-classical states of light such as squeezed or entangled states \cite{{Kimble1987}, {Grangier1987}, {Yurke1986}, {Scully1993}}. The introduction of squeezed vaccuum for sub-shot noise phase detection is now implemented in gravitational wave detection experiments \cite{{Schnabel2010}, {Grote2013}}. Squeezing enhanced optical phase tracking for optical comminication applications has also been demonstrated \cite{Yonezawa2012}. Another class of measurements using adaptive feedback mechanism have been suggested for achieving accuracy below the SNL \cite{Wiseman1997, Wiseman2002, Wiseman2006}. In the context of optimally estimating a classical Markov process that is coupled to a quantum sensing system a time symmetric quantum smoothing framework has been developed and demonstrated experimentally \cite{{Tsang2009}, {Wheatley2010}}. Fundamental quantum limits to time-varying waveform detection have been discussed recently in the context of force estimation problem \cite{{Tsang2011}, {Tsang2012}}. A stochastic Heisenberg limit has also been studied in the context of optimally estimating time-varying fluctuating phase \cite{Wiseman2013}. In the present work we use a Mach-Zehnder interferometer setup without any additional hardware and illustrate enhanced phase detection accuracy based on the redundancy/sparsity of the phase function to be measured.  

The analysis that leads to SNL is traditionally performed for point detectors and a phase extraction procedure such as balanced homodyning or phase shifting is assumed. This leads to all the data points in time domain (e.g. photon counts recorded by a point detector as a function of time) or in space domain (e.g. pixels of an array detector) being processed in parallel. In most practical applications the underlying solution $\phi_O$ that one is seeking has some structure (as opposed to random or white noise)  and hence the individual measurement points in the interference data may not be treated as independent of each other. Recent developments in the area of compressive sensing \cite{Candes2006} suggest that such redundancy in the desired solution may be exploited to achieve excellent signal/image recovery even with data that is traditionally considered incomplete. This expected redundancy in the signal to be recovered is not considered in methods such as PSM but can be modelled in an optimization framework to gain noise advantage as we illustrate here. 

For the phase measurement problem we consider a constrained optimization formulation \cite{Bertero1998} where we minimize a cost function of the form \cite{Khare2013}:
\begin{eqnarray}\label{eq:CO}
C(O, O^{\ast}) = || \beta(I) &[I - ( |R|^2 + |O|^2 + R^{\ast}O + R O^{\ast} )] ||^2 \nonumber \\&+ \alpha \psi(O, O^{\ast}).
\end{eqnarray}
The first term in the above equation is a weighted L2-norm squared data fit and the second term is a constraint that models some physically desirable property of the solution $O$. The choice of $\psi(O,O^{\ast})$ depends on the problem at hand as we shall explain later. The weights $\beta(I)$ in the first term may be selected such that the measurements with larger photon counts get more importance in the cost function. The parameter $\alpha$ controls the relative importance of the two terms in the cost function. The knowledge of $R$ is required for both the phase shifting and the constrained optimization methods in order to determine the amplitude and phase of $O$. Recently we have demonstrated the advantage of such an approach for achieving single shot high resolution digital holographic imaging \cite{{Khare2013}, {Khare2014}}. These experiments were however performed at high light level and the issues such as accuracy relative to SNL were of no concern there as is the case in the present study with low photon level interference data. 
\begin{figure}[tb]
\begin{center}
\includegraphics[width=3in]{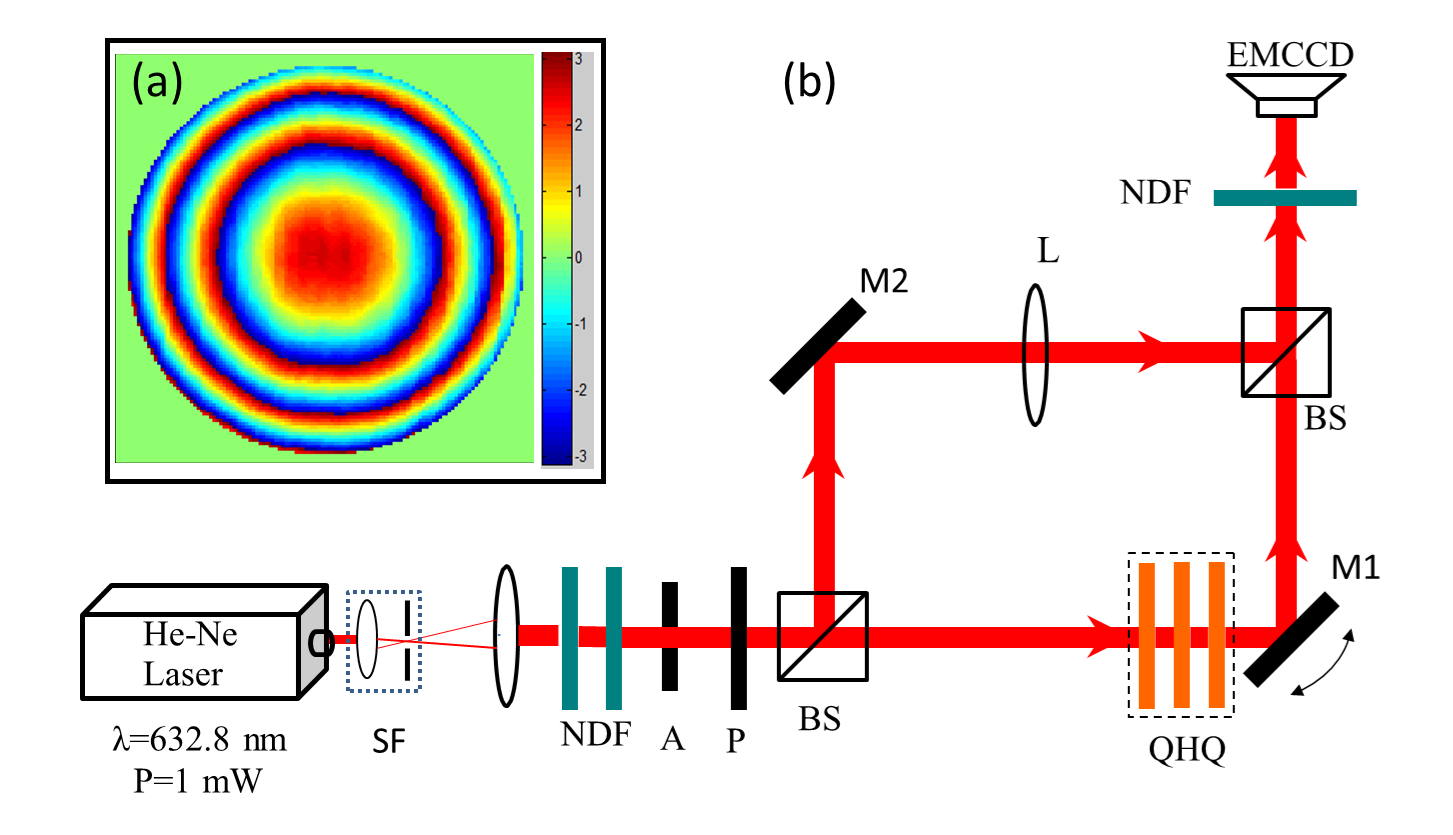}
\caption{(a) Reference phase map at the sensor plane obtained using high light level ($> 5000 $ counts/ pixel) phase shifting data. (b) Experimental setup for low light level interference experiment. SF: Spatial filter, P: polarizer, A: aperture, NDF: Neutral density filter, QHQ: Geometric phase shifter , L: lens ($f = 10$ cm), BS: Beamsplitter, M1, M2: Mirrors, EMCCD: Electron Multiplier CCD sensor. }
\label{fig:setup}
\end{center}
\end{figure}
In order to test the noise characteristics of this optimization approach to phase detection, we performed a low light level interference experiment where a tilted plane wavefront and a quadratic wavefront were interfered. In order to obtain data that is photon noise limited, we employed a sensitive EMCCD array sensor (Make: Andor iXon3) in photon counting mode. A $128 \times 128$ pixel region of the EMCCD was used for all the illustrations below.The schematic setup of our experiment as shown in Fig. \ref{fig:setup} consists of a Mach-Zehnder interferometer. The illumination source is a linearly polarized He-Ne laser which is collimated and split at the first beamsplitter. The mirror M1 in the reference arm is used to produce a tilt in the plane reference wavefront. The lens L ($f = 10$ cm) in the object arm produces an approximately quadratic phase front. The QHQ (Q = quarter wave plate, H = half wave plate) arrangement in the reference arm was used as a geometric phase shifter \cite{Mukunda1993} for generating four frames of the phase shifting interference data . The optimization procedure as in Eq. (\ref{eq:CO}) requires a single interference data frame. A separate interference data frame with number of photon counts approximately equal to the sum of photon counts in the four phase shifting frames was thus recorded. This single interference frame was then used with the optimization algorithm. The performance of the phase estimation methods is compared against the average number $N_0$ of photon counts registered per EMCCD pixel. In our experimental tests $N_0$ varied from approximately 800 to 10.  Light level reduction may be achieved with the help of neutral density filters or by controlling the exposure time (electronic shutter) of the EMCCD array. 
\begin{figure}[tb]
\begin{center}
\includegraphics[width=3in]{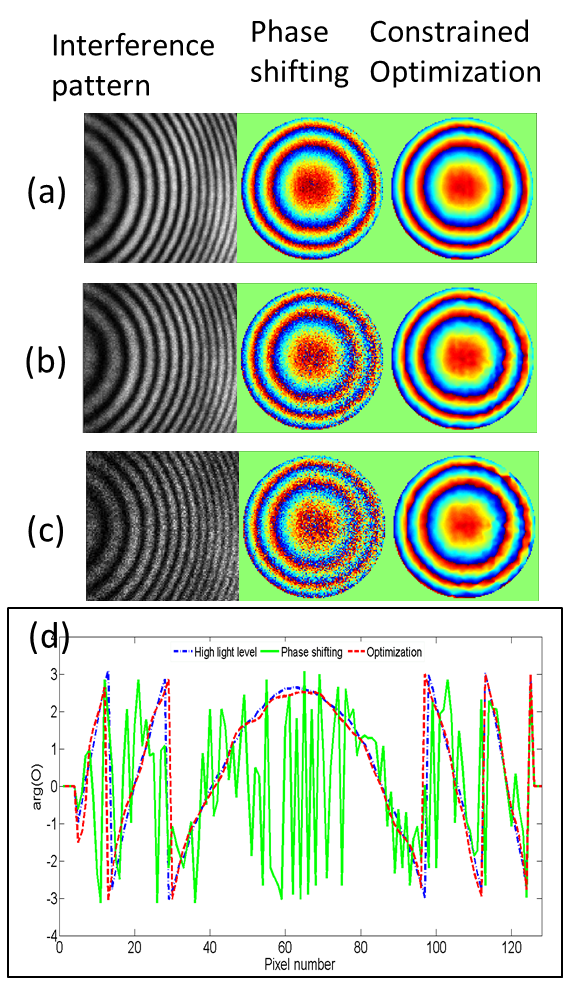}
\caption{(a)-(c) Single shot interference patterns used in the optimization method (left column), phase recovery using the PSM (middle column) and the optimization method (right column). The $N_0$ values in (a) - (c) are 225, 58, 10 respectively.  (d) Phase profiles of the center pixel column of the image $\phi_{HLL}$(Fig. \ref{fig:setup}(a)) and the phase recoveries using PSM and optimization methods as in (c) above for $N_0$ = 10 counts/pixel. }
\label{fig:PS_CO}
\end{center}
\end{figure}
In order to find a ground truth phase map of the object beam $O$ for a later comparison with the low light level phase reconstructions, a phase shifting dataset with sufficiently high light level ($N_0 > 5000$) was recorded. This high light level dataset is able to give a smooth phase map $\phi_{HLL}$ for the object beam as shown in Fig. \ref{fig:setup}(a). In order to find $R$, a separate calibration interference pattern was recorded without any object in the object arm of the interferometer and the straight line fringes were used to estimate the tilt in the reference beam. The constrained optimization procedure was implemented using a gradient descent iteration. Since the cost function is a function of both $O$ and $O^{\ast}$ the steepest descent direction is computed with respect to $O^{\ast}$ \cite{Brandwood1983}. The gradient of the cost function in Eq. (\ref{eq:CO}) is given by:
\begin{eqnarray}
&&\nabla_{O^{\ast}}C(O, O^{\ast}) \nonumber\\&&= - \beta(I)[I - ( |R|^2 + |O|^2 + R^{\ast}O + R O^{\ast} )] (O + R) \nonumber\\&& \;\;\;\; + \alpha \nabla_{O^{\ast}} \psi(O, O^{\ast}). 
\end{eqnarray} 
The iterative algorithm is then designed such that an updated solution $O^{(n+1)}$ is obtained from the previous solution $O^{(n)}$ as:
\begin{equation}
O^{(n+1)} = O^{(n)} - t [\nabla_{O^{\ast}}C(O, O^{\ast})]_{O = O^{(n)}}.
\end{equation}          
The step size $t$ may be selected in each iteration by standard backtracking line search \cite{BoydCV}. We used the weights $\beta(I)$ with $I$ in photon count units proportional to $\sqrt{I}$ so that the terms with higher photon counts were weighted by their relative detection signal-to-noise ratio. Further, since the object wavefront has resulted due to Fresnel diffraction from the (lens) object, the resultant field $O$ is expected to have certain degree of smoothness. The smoothness property for Fresnel diffraction field is expected irrespective of any sharp features that the object may have. This desirable property can be modelled with the penalty term $\psi(O, O^{\ast})$ defined as:
\begin{equation}
\psi(O, O^{\ast}) = \sum_p \sum_{q \in N_p} w_{pq} |O_p - O_q|^2.
\end{equation}
The first summation above is over all pixels $p$ in the image and the pixels $q$ belong to some neighborhood $N_p$ of a particular pixel $p$. The window function $w_{pq}$ is a decreasing function (e.g. a Gaussian) of the distance between the pixels indexed by $p$ and $q$. From the nature of $\psi(O, O^{\ast})$ it may be noted that large differences in the numerical value of $O$ at any pixel with those in its neighborhood are penalized and a locally smooth solution as guided by window function $w_{pq}$ is obtained.   In practice we implemented the optimization algorithm by alternatingly minimizing the two terms of the cost function in an adaptive manner in a fashion similar to some recent work in image recovery literature \cite{{Pan2008}, {Ritschl2011}}. 

\begin{figure}[tb]
\begin{center}
\includegraphics[width=3.25in]{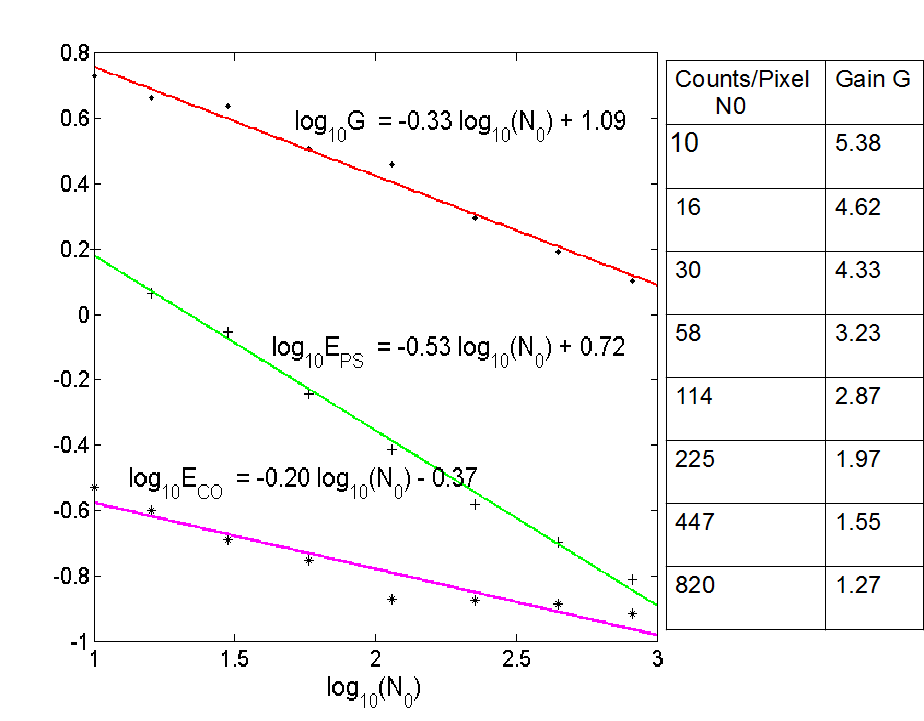}
\caption{Plot of noise gain $G$, $E_{PS}$, $E_{CO}$ (Eq. (\ref{eq:gain})) with respect to average number $N_0$ of photon counts per pixel (on log-log scale) used for phase map estimation.}
\label{fig:GlogN0}
\end{center}
\end{figure}

For our experimental data, approximately 15-20 iterations were required in each case for achieving the convergence. The relative change in the solutions from successive iterations was seen to be less than $10^{-3}$ (or 0.1 \%) at this stage. Some of the phase recovery results are shown in Fig. \ref{fig:PS_CO} (a)-(c). The phase maps for the object field as obtained using the PSM (Eq. \ref{eq:PS}) and the corresponding result using the constrained optimization method are shown such that both the methods use the same average number of photons per pixel. We clearly observe the advantage of using the constrained optimization procedure by visual comparison of the resultant phase maps with the phase map $\phi_{HLL}$ as in Fig. \ref{fig:setup}(a). Denoting the phase maps obtained using the PSM and the constrained optimization approaches as $\phi_{PS}$ and $\phi_{CO}$ respectively, we define the noise gain as:
\begin{equation}\label{eq:gain}
G = \frac{E_{PS}}{E_{CO}} = \frac{||\phi_{HLL} - \phi_{PS}||}{||\phi_{HLL}-\phi_{CO}||}.
\end{equation} 
The gain $G$ is a ratio of the RMS (or L2-norm) phase errors in $\phi_{PS}$ and $\phi_{CO}$ with respect to $\phi_{HLL}$ (Fig. \ref{fig:setup}(a)). In Fig. \ref{fig:GlogN0}, we show log-log plots of the gain G, and the two RMS errors $E_{PS}$, $E_{CO}$ as in Eq. (\ref{eq:gain}) with respect to $N_0$. $E_{PS}$ is observed to scale as $N_{0}^{-0.53 \pm 0.04}$ which is close to the expected shot noise behaviour, whereas $E_{CO}$ is seen to scale as $N_{0}^{-0.20 \pm 0.06}$. The noise gain $G$ is seen to scale as $N_{0}^{-0.33 \pm 0.04}$. Here the $\pm$ ranges in the scaling relations show 95\% confidence interval for the scaling coefficient for fitting of our data. While we have made experimental measurements for $N_0$ as low as 10 based on detector limitations, our tests on simulated interference patterns for lower photon counts (up to $N_0 = 1$) show that the trend in scaling of $E_{PS}, E_{CO}$ and $G$ as above continues to hold. An RMS error scaling of $N_0^{-0.25}$ has been obtained in a feedback based interferometric scheme in \cite{Wiseman2002} for time-varying phase signals. The scaling law obtained by us is however likely to change depending on the sparsity in the phase function to be measured. It is more important to note from Fig. \ref{fig:GlogN0} that in the range of $N_0$ considered the error $E_{CO}$ for the optimization method is is always lower than the error $E_{PS}$ for the phase shifting method. Since the two solutions $\phi_{PS}$ and $\phi_{CO}$ are almost equal at high light levels, the optimization solution is significantly better as $N_0$ is reduced. For example, if $N_0$ is reduced by a factor of 2, the PSM solution gets worse by $\approx \sqrt{2}$ whereas the optimization solution gets worse by a factor $2^{0.20} = 1.15$.  The weak dependence of $E_{CO}$ on $N_0$ in the low $N_0$ range highlights the importance of the smoothness penalty term in the optimization solution.  A further analysis leading to a generalized multipixel SNL is required that incorporates the statistics of the light source as well as a measure of redundancy in the phase function $\phi_O$ that is to be estimated. 

The noise gain $G$ may be interpreted in two different ways- when traditional approaches such as PSM are used and ideal detectors are assumed, achieving the similar accuracy as offered by the optimization method will require: (i) classical light that is more intense by a factor of $G^2$ or (ii) non-classical sub-Poissonian light with fluctuations below the shot noise by a factor of $G$. In our opinion, the noise gains $>$ 5 as observed in our experiments can be significant for sensitive phase detection applications that are currently considered limited by shot noise. The optimization framework we have used here exploits the redundancy in the function $\phi_O$ to achieve improved phase detection accuracy even when a classical light source is used. We expect further improvement in phase detection accuracy if non-classical states of light (e.g. squeezed states, spatially entangled light field) or schemes such as adaptive feedback are is used in combination with this unconventional optimization based approach to phase estimation. While we have considered a stationary 2D wavefront in this work, a similar approach will apply equally well if a series of interference data points is recorded in time with a point detector and an appropriate penalty term is designed that models the desirable properties of a time varying phase function. Also we are not restricted to the smoothness penalty function as used in this work  - other forms of penalties such as L1-norm based penalties (e.g. Total Variation) or generalized Gibbs priors \cite{Geman1984} may well be used if required. 

In conclusion, our work suggests that noise performance better than conventional single pixel based SNL for phase detection in an interference experiment may be achievable even with classical light if an optimization approach to phase detection as described here is used. Any interferometric scheme (using either classical or non-classical states of light) is expected to benefit from such an approach to achieve enhanced phase detection sensitivity. The limits such as SNL that are traditionally defined with considerations on statistics of the light source alone may thus be generalized to take into account the redundancy in the phase signal that we intend to measure. The authors acknowledge discussions with Dr. V. Ravishankar and Dr. M. S. Santhanam. MS and KK acknowledge support from DBT India grant BT/PR8008.

\end{document}